# STUDY OF RECONFIGURABLE MOSTLY DIGITAL RADIO FOR MANET

**Aubin Lecointre\*, Daniela Dragomirescu\*, Robert Plana\***
\* University of Toulouse
LAAS-CNRS
7, avenue du Colonel Roche
31077 Cedex 4, FRANCE
{alecoint, daniela, plana}@laas.fr

*Abstract*
*We introduce the radio reconfigurability thanks to IR-UWB mostly digital architecture for MANET context. This particular context implies some constraints on the radio interface such as low cost, low power, small dimensions and simplicity. Here, we propose an implementation of dynamic reconfigurable receiver on ASIC, and FPGA, after having explained the advantages of mostly digital radio for reconfigurability. In this paper, by studying our prototypes, we could prove that reconfigurability is on the contrary with MANET constraints needs. The proposed solution allows data rate, radio range, energy and spectrum occupation reconfigurability.*

## 1. INTRODUCTION

This paper addresses the MANET (Mobil Ad hoc NETwork) problematic according to the radio interface point of view. In this context there are a lot of applications, such as sensors networks, home automation, military applications, entertainment network, etc … Our goal is to propose a reconfigurable transceiver architecture for dealing with the whole of MANET applications and theirs distinct needs. That is to say, we try to introduce a unique radio interface which is able to respond to any kind of MANET radio interface constraints. We will use IR-UWB (Impulse Radio Ultra WideBand) [1] since it could be designed as a mostly digital radio [2], so implying a good capacity for reconfigurability. In addition IR-UWB is a viable solution for MANET constraints, which are: low cost, low power, small dimensions and simplicity. While IR-UWB deals with these four constraints, reconfigurability has to answer to applications constraints (data rate, radio range, spectrum consideration, etc …)

The paper is laid out as follow. In Section 2, we will remind briefly IR-UWB mostly digital radio concept. Section 3 allows exposing the reconfigurability principle. The transceiver hardware implementation in ASIC (Application Specific Integrated Circuit) or FPGA (Field Programmable Gate Array) will be explored in Section 4. Before conclusion, in the Section 5 we will compare ours implementation according to the MANET constraints.

## 2. IR-UWB PRINCIPLE, A MOSTLY DIGITAL RADIO

IR-UWB is defined as a radio technique which uses more than 500 MHz of 10 dB bandwidth, by using very short impulse. Pulse modulations such as PPM (Pulse Position Modulation), OOK (On Off Keying) or BPAM (Binary Pulse Amplitude Modulation) are used for transmitting information [1]. Besides Time-Hopping (TH) [3] is also used for multi users capability and because it randomizes the radio signal. Thanks to its time domain approach IR-UWB emitter could be designed without RF stage, including mixer, VCO (Voltage Controlled Oscillator), etc … (cf. figure 1) [2] [4].

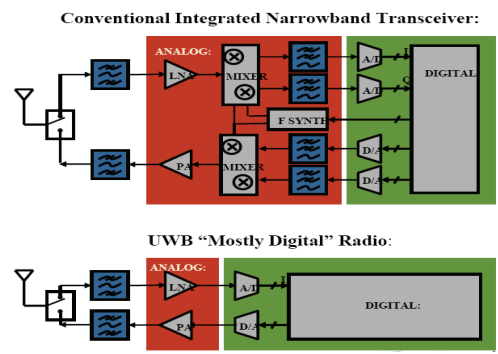

*Fig. 1. Illustration of the mostly digital radio concept.*

Indeed IR-UWB emitter consists in modulating a very short pulse and applying it a delay for TH. By implementing the pulse waveform in RAM (Random Access Memory), we could easily set up an IR-UWB receiver. For example for OOK modulation we have to use this saved pulse waveform when we want to send a binary one. The concern with this technique is dimensioning. Indeed, the DAC (Digital to Analog Converter) is the key point of the emitter architecture. The characteristics of this latter will defined the capabilities of the emitter. Its clock rate and its bit resolution will determine the achievable performance of our emitter. For example the pulse bandwidth depends on these two parameters. The bandwidth is linked to the clock rate, i.e. the sampled frequency, with the Shannon theorem [5]. Sub-sampling theory could be impact in a lesser way the performances of the UWB system. Indeed, in digital system, the sampling Shannon rate has to be twice higher than the maximum frequency. In classical narrow band technique, there is no difficulty, but with IR-UWB and its very large bandwidth, it's problematic to obtain such sampling frequency. Since the IR-UWB band is equal to several gigahertz. Sub-sampling theorem indicates that a sampling rate twice higher than the bandwidth used is enough under some conditions. Here we propose to use this theorem for obtaining mostly digital radio.

Since emitter and receiver are mostly digital, they could be easily implement on ASIC or FPGA. Thus the transceiver is composed of a ADC/DAC and a digital part, either a FPGA or an ASIC. By using a mostly digital architecture it is possible to implement reconfigurable radio.

## 3. IR-UWB RECONFIGURABLE RADIO

Inspired from the software defined radio concept, reconfigurability consists in provide a transceiver able to adapt its ability to application needs. For example the transceiver could change the data rate, the radio range or spectrum occupation.

There are two reconfigurability visions, described in figure 2, the first one profit from FPGA configuration for changing the architecture of the receiver, while the second one do not use FPGA capacity and base its concept on reconfigurable parameters. Thus, the first one is reconfigurable architecture, while the second one is reconfigurable parameters. The second solution could be used with ASIC or FPGA.

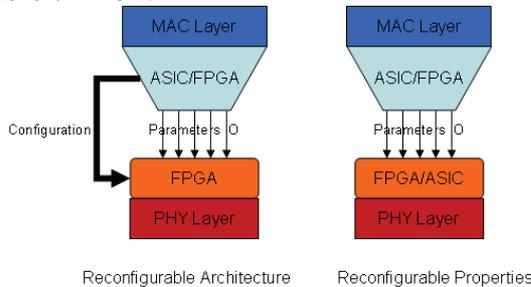

*Fig. 2. Two solutions of reconfigurability concept.*

The reconfigurable architecture use configuration FPGA capability for changing the transceiver architecture each time a reconfigurability process is started. The pilot component which is in charge of reconfiguring the FPGA could dispose in its memory different versions of FPGA architecture and set up them when it is asked. Or the pilot component, which could be a FPGA or an ASIC, could have a large range of bloc unit and when the reconfiguration is started, it have to assemble them in order to obtain a new transceiver with new abilities (cf. figure 3).

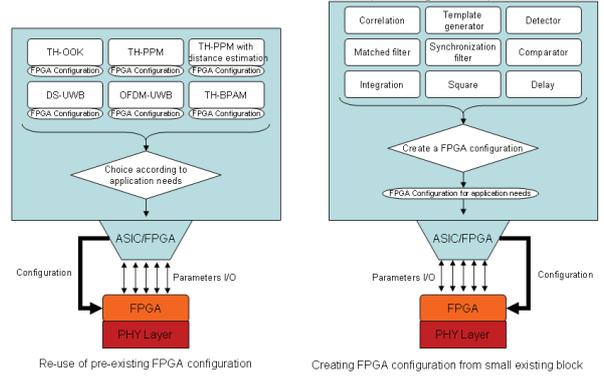

*Fig. 3. Two solutions for dynamic reconfigurability.*

We implement the reconfigurable parameters circuits. In order to obtain reconfigurable data rate, radio range and spectrum occupation, we have to parameterize the IR-UWB transceiver. The reconfiguration process does not change the architecture but change the key values of the transceiver, such as TH parameters (frame duration, time slot duration, and the number of time slot per frame), IR-UWB parameters (pulse duration, pulse amplitude, and pulse waveform).

If we compare these two solutions we can see that they have both advantages and drawbacks. Concerning the respect of MANET constraints, the properties reconfigurable vision is better, since only one ASIC is used for the PHY layer, while the architecture vision proposes to use a FPGA and an ASIC. The problem when we use FPGA, is whatever the circuit implemented on it, we will have the performance of the FPGA model. Nevertheless using the architecture reconfigurable solution allows obtaining a larger achievable performance panel.

In this paper we will only consider the reconfigurable parameters vision. With IR-UWB, we could express the data rate as follow:

$$D_{total}(bits/s) = \frac{N_c}{T_f} = \frac{N_c}{Nc \times T_c} = \frac{1}{T_c}$$

With Tc, Tf, and Nc, respectively, the time slot duration, the frame duration, and the number of time slot per frame. As a result, changing one of these key values imply a data rate modification. Thus, for having a reconfigurable transceiver, we have to change this parameters by means of a reconfigurability process.

Concerning the spectrum occupation reconfigurability, with IR-UWB, it depends on the pulse waveform, pulse amplitude, and duration amplitude. The figure 4 exposes a panel of distinct non-modulated pulse and their spectrum occupation with different configurations: waveform/amplitude/duration.

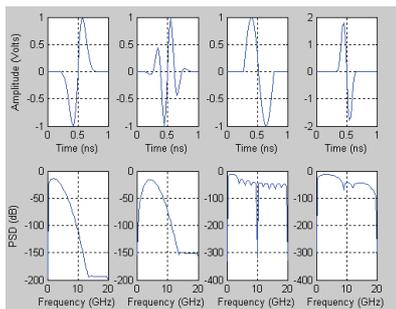

*Fig. 4. UWB pulse waveforms and their spectrum.*

We have to balance two criteria: the radio range and the energy consumption. They will depend on a lot of parameters such as pulse amplitude, spectrum occupation, the sample frequency and the ASIC/FPGA operating frequency. For example, at the same level of puissance, using a lower frequency band permits to achieve a better radio range than using a high frequency band.

Before exposing how we have set up the reconfigurability concept at the hardware implementation level, the figure 5 illustrates our high level reconfigurability vision.

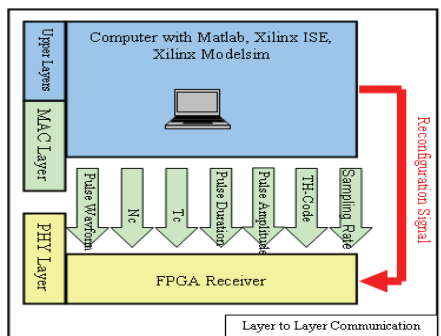

*Fig. 5. High level reconfigurability illustration for dynamic properties reconfigurable solution.*

Figure 5 presents our reconfigurable architecture for setting up dynamic reconfigurability. Once the reconfiguration signal is send from the MAC (Medium Access Control) layer to the PHYsical layer, the transceiver is able to change its properties while the system is in used.

## 4. OUR SOLUTION FOR AN IR-UWB RECONFIGURABLE TRANSCEIVER

Since we have chosen the second reconfigurability vision, we have to set up a reconfigurable parameters receiver on FPGA or ASIC. In order to achieve it, we have implemented reconfigurable parameters as entries at the interface between the MAC (Medium Access Control) and the PHYsical layer (cf. figure 5).

Our digital circuits are developed thanks to VHDL (VHSIC Hardware Description Language) language. Thus reconfigurable parameters are defined as follow in order to be embodied as entries:

```
entity reconfigurable_receiver is
Port (
-- inputs
        CLK : in std_logic;
        RESET : in std_logic;
        Renable : in std_logic;
        signal_recu : in std_logic_vector(31 downto 0);
        load_code : in std_logic;
        lg_code : in integer range 0 to 255;
        unload_code :  in std_logic;
        code_j_data : in integer range 0 to 255;
-- reconfigurable parameters
        nb_Tc_par_trame_TH : in integer range 0 to 255;
        Tc : in integer range 0 to 255;
-- reconfiguration signal
        sig_reconf : in std_logic;
-- ouputs
        out_recepteur_trame : out std_logic;
        out_recepteur_chip : out std_logic;
        rythme_out_recepteur_chip : out std_logic;
        rythme_out_recepteur_trame : out std_logic
);
end  reconfigurable_receiver;
```

Figure 6 exposes the implementations of these reconfigurable parameters by means of a RTL schematic view of one IR-UWB transceiver.

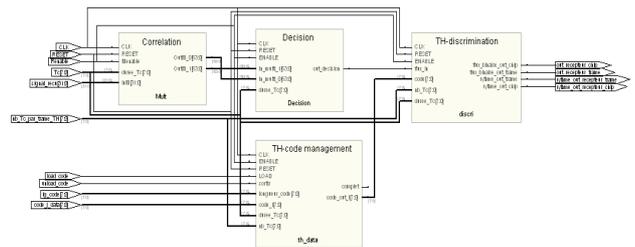

*Fig. 6. RTL schematic view of an IR-UWB data rate and TH-code reconfigurable receiver.*

Nevertheless this technique has some limitations due to the limit size of the VHDL entity entries. Entry size will define the variation amplitude of the value of the different parameters (Time slot duration, number of time slot per frame, pulse amplitude, pulse waveform, etc. …). Indeed it determines the achievable data rate, spectrum occupation, radio range, and the others reconfigurable receiver properties. In other word, we have a dimensioning concern. Achievable performances are linked to the entry size.

For comparing, we use three solutions: ASIC implementation with a design kit: AMS 0,35 µm and Xilinx FPGA: a Spartan III and a Virtex 5.

## 5. IMPLEMENTATION AND COMPARATIVE ANALYSIS

During our study, we will compare two versions of the reference implementation: a TH-PPM IR-UWB receiver. One version is static, while the second one has the dynamic reconfigurability. For this experimentation, we will study only data rate and TH-code reconfigurability.

These two implementations are evaluated according MANET context criteria: size, energy consumption. We have also had two performances criteria: data rate, which depends on frequency and the BER versus SNR. Concerning the BER (Bit Error Rate) versus SNR (Signal to Noise Ratio), the two solutions are identical.

Table I, II and III compare them for each kind of target: ASIC, Spartan III, and Virtex 5.

*Table I. Impact of reconfigurability on ASIC implementation*

| ASIC Synthese Report Design AMS 3.60 at 0.35µm | | | | |
|---|---|---|---|---|
| VERSION | Frequency | Clock | Total Area | Dynamic Power (mW) |
| Static TH-PPM receiver | 333 MHz | 3 ns | 152123.937500 | 23.8080 mW |
| Reconfigurable TH-PPM receiver | 50 MHz | 20 ns | 1254628.500000 | 18.8108 mW |

*Table II. Impact of reconfigurability on FPGA Spartan III implementation*

| Xilinx FPGA Spartan III | | | |
|---|---|---|---|
| | Synplicity Synthesis | Placement and Routage by Xilinx | |
| VERSION | Frequency | Frequency | Size (in gates) |
| Static TH-PPM receiver | 160,8 MHz | 129,416 MHz | 6466 |
| Reconfigurable TH-PPM receiver | 84,9 MHz | 62,672 MHz | 55054 |

*Table III. Impact of reconfigurability on FPGA Virtex 5 implementation*

| Xilinx FPGA Virtex 5 | | | |
|---|---|---|---|
| | Synplicity Synthesis | Placement and Routage by Xilinx | |
| VERSION | Frequency | Frequency | Size (in gates) |
| Static TH-PPM receiver | 448,7 MHz | 382,117 MHz | 6232 |
| Reconfigurable TH-PPM receiver | 128,9 MHz | 104,3 MHz | 15422 |

These tables give us an idea of the impact of the reconfigurability on receiver performances. By analyzing table I, II and III, we could say that reconfigurability implies a decrease of data rate (data rate is linked to frequency) and an increase of circuit size on both ASIC and FPGA implementation, which was expected.

We have to make the same analysis for energy consumption criteria. On the FPGA we'll have the dynamic power due to circuit commutation and also a static power necessary to maintain the FPGA configuration. The dynamic power will depend of the number of gates like in the ASIC implementation. The results on size and dynamic power criteria on ASIC implementation prove that reconfigurability has a cost in the MANET context. Indeed, the complexity is higher (and the maximum frequency is lower), the data rate is lower, and the dimensions are bigger, but we can target more application with the same circuit.

We can also note that using Virtex 5 instead of Spartan III allow reaching higher data rate. However, a Virtex 5 is more expensive.

As a result, our study has exposed how we could set up dynamic reconfigurability. Moreover, we have estimated the cost of this essential capacity in the MANET context. The FPGA solution is used for prototyping the circuit, but the final circuit will be on ASIC because of its incontestable advantages in MANET context: low size and low power. We can suppose that low cost circuits could be done because of a high production volume thanks to very demanding targeted application and the use of parameter reconfigurable approach.

## 6. CONCLUSION

In this paper we have exposed a reconfigurable transceiver based on IR-UWB and its mostly digital architecture capacity. Our reconfigurability vision permits to adapt the IR-UWB signal according to four characteristics: radio range, data rate, spectrum occupation and energy consumption. We have proposed a reconfigurable receiver implementation, based on parameterization by means of entries, on three distinct targets (ASIC AMS 3.60 0.35um, Spartan III, Virtex 5). These three prototypes are used for determining the reconfigurability impact and cost on receiver implementation in the MANET context. Dynamic reconfigurability implies additional cost, an increase of size, and a loss in term of data rate.

In perspective, we work on this reconfigurable receiver design using a very advance technology from ST Microelectronics, the Silicon CMOS SOI 65 nm to obtain a really low cost, low size and low power, completely integrated IR-UWB receiver at 60 GHz.

## 7. REFERENCES

[1] I. Oppermann, et al., "UWB Theory and Applications", Wiley 2004.
[2] Ian O'Donnell, et al., "An Integrated, Low-Power, Ultra-Wideband Transceiver Architecture for Low-Rate Indoor Wireless Systems", Berkeley Wireless Research Center, Univ. of California, Berkeley, IEEE CAS Workshop on Wireless Communications and Networking, Pasadena, 2002.
[3] M.Z. Win, R.A. Scholtz (1998), Impulse radio: how it works, IEEE Communications Letters, vol. 2, no. 2.
[4] D. Morche, et al., "Vue d'ensemble des architectures RF pour l'UWB", LETI, UWB Summer School, France 2006.
[5] Mike Shuo-Wei CHEN, et al., "A subsampling UWB Impulse Radio Architecture Utilizing analytic signalling", IEICE TRANS. ELECTRON. June 2006.